\documentclass{iaus}
\usepackage{graphicx}

\title[Binaries and the dynamical mass of star clusters] 
{Binaries and the dynamical mass of star clusters}

\author[M.B.N. Kouwenhoven \& R. de Grijs]   
       {M.B.N. Kouwenhoven$^1$ 
	 \and
	 R. de Grijs$^{1,2}$}

\affiliation{$^1$Department of Physics and Astronomy, University of Sheffield, \break Hicks Building, Hounsfield Road, Sheffield S3\,7RH, United Kingdom \break email: t.kouwenhoven@sheffield.ac.uk\\[\affilskip]
$^2$ National Astronomical Observatories, Chinese Academy of Sciences, \break 20A Datun Road, Chaoyang District, Beijing 100012, P.R.~China \break email: r.degrijs@sheffield.ac.uk}

\pubyear{2007}
\volume{264}  
\pagerange{119--126}
\date{?? and in revised form ??}
\jname{Proceedings Title IAU Symposium}
\editors{A.C. Editor, B.D. Editor \& C.E. Editor, eds.}
\begin{document}

\maketitle

\begin{abstract}
The total mass of a distant star cluster is often derived from the virial theorem, using line-of-sight velocity dispersion measurements and half-light radii, under the implicit assumption that all stars are single (although it is {\em known} that most stars form part of binary systems). The components of binary stars exhibit orbital motion, which increases the measured velocity dispersion, resulting in a dynamical mass overestimation. In these proceedings we quantify the effect of neglecting the binary population on the derivation of the dynamical mass of a star cluster. We find that the presence of binaries plays an important role for clusters with total mass $M_{\rm cl} \leqslant 10^5~{\rm M}_\odot$; the dynamical mass can be significantly overestimated (by a factor of two or more). For the more massive clusters, with $M_{\rm cl} \geqslant 10^5~{\rm M}_\odot$, binaries do not affect the dynamical mass estimation significantly, provided that the cluster is significantly compact (half-mass radius $\leqslant 5$~pc). 
\keywords{Star clusters, binaries, numerical methods}
\end{abstract}

\firstsection 
\section{Introduction}

Young star clusters, with typical masses of $M_{\rm cl}=10^{3-6}~{\rm M}_\odot$, indicate recent or ongoing violent star formation, and are often triggered by mergers and close encounters between galaxies. Only a fraction of these young massive star clusters evolve into old globular clusters, while the majority ($60-90\%$) will dissolve into the field star population within about 30\,Myr (e.g., de Grijs \& Parmentier 2007). In order to understand the formation and fate of these clusters, it is important to study these in detail, and obtain good estimates of the mass, stellar content, dynamics, and binary population. 
The dynamical mass for a cluster in virial equilibrium, consisting of single, equal-mass stars, is given by:
\begin{equation}
M_{\rm dyn} = \eta \, \frac{ R_{\rm hm} \sigma_{\rm los}^2 }{G}
\end{equation}
(Spitzer 1987), where $R_{\rm hm}$ is the (projected) half-mass radius, $\sigma_{\rm los}$ the measured line-of-sight velocity dispersion, and $\eta \approx 9.75$. For unresolved clusters, $\sigma_{\rm los}$ is usually derived from spectral-line analysis, neglecting the presence of binaries. However, observations have shown that the majority of stars form in binary or multiple systems (e.g., Duquennoy \& Mayor 1991; Kouwenhoven et al. 2005, 2007; Kobulnicky et al. 2007). When binaries are present, $\sigma_{\rm los}$ does not only include the motion of the binaries (i.e., their centre-of-mass) in the cluster potential, but additionally the velocity component of the orbital motion. This results in an overestimation of the velocity dispersion, and hence of $M_{\rm dyn}$.

\section{The effect of binaries on the dynamical mass $M_{\rm dyn}$ of a star cluster}

The systematic error introduced by the single-star assumption depends on the properties of the star cluster and of the binary population. This effect is most easily seen in the extreme case when $\sigma_{\rm los}$ is dominated by the orbital motion of binaries (which is the case for most Galactic OB associations). In this binary-dominated case, the measured $\sigma_{\rm los}$ is independent of $R_{\rm hm}$ and the true cluster mass, $M_{\rm cl}$. The inferred $M_{\rm dyn}$ from equation~(1) is then proportional to $R_{\rm hm}$. The dynamical mass overestimation is therefore $M_{\rm dyn}/M_{\rm cl} \propto R_{\rm hm}/M_{\rm cl}$. Sparse clusters are thus most sensitive to binaries; the systematic error in $M_{\rm dyn}$ could be as influential as that from the assumption of virial equilibrium. The cluster structure, stellar mass function and the presence of mass segregation also affect $M_{\rm dyn}$, but these are of much less importance (Kouwenhoven \& de Grijs, in prep.). 

Whether or not the binaries affect $M_{\rm dyn}$ additionally depends on the properties of the binary population. Obviously, the most important parameters are the binary fraction (which determines the relative weight between singles and binaries when measuring $\sigma_{\rm los}$) and the semi-major axis, $a$, or period distribution (tight binaries have a larger orbital velocity component). As in a binary orbit $v_{\rm orb} \propto a^{-1/2}$, we have for the binary-dominated case $M_{\rm dyn} \propto \sigma_{\rm los}^2 \propto a^{-1}$. The distributions over eccentricity and mass ratio play a significantly smaller role than the binary fraction and semi-major axis distribution (Kouwenhoven \& de Grijs, in prep.).

Finally, we wish to stress that further systematic errors are introduced by observational selection effects. Firstly, $\sigma_{\rm los}$ is often derived from spectral lines of red giants; the velocities of these objects may or may not be representative for the cluster as a whole. Secondly, a measurement in the cluster centre, and near the tidal limit, will result in dynamical masses differing by $\sim 50\%$; caution should be exercised when interpreting the observational results. 

\section{When is binarity important?}

The effect of binaries on the dynamical mass determination of a star cluster is important if the typical orbital velocity of a binary component is of order, or larger than, the velocity dispersion of the particles (single/binary) in the potential of the cluster. Our simulations indicate that, for example, the dynamical mass is overestimated by 70\% for $\sigma_{\rm los} = 1$\,km\,s$^{-1}$, 50\% for 2\,km\,s$^{-1}$, 20\% for 5\,km\,s$^{-1}$, and 5\% for 10\,km\,s$^{-1}$. Due to spectral resolution and stellar atmospheric turbulence, most {\em measured} velocity dispersions are $\sigma_{\rm los} \geqslant 5$\,km\,s$^{-1}$. Most of the known dynamical masses of massive star clusters are therefore only mildly affected by the presence of binaries. However, for low-mass star clusters, the spectroscopic velocity dispersion may result in an overestimation of the dynamical mass by a factor of two or more.


\end{document}